\RequirePackage{lineno}
\documentclass[preprint,aps,prc,showpacs,superscriptaddress,preprintnumbers,floatfix,nofootinbib]{revtex4}
\usepackage{epsfig,graphics}
\usepackage{graphicx}% Include figure files
\usepackage{dcolumn}% Align table columns on decimal point
\usepackage{bm}% bold math
\usepackage{amsmath}
\usepackage[usenames]{color}
\usepackage{ulem} %% for strike-through

\newcommand{\lr}[1]{\left\langle #1\right\rangle}

\newcommand{\comments}[1]{}

\begin{document}

\title{Multiparticle azimuthal cumulants in p+Pb collisions from a multiphase transport model}

\author{Mao-Wu Nie}
\affiliation{Shanghai Institute of Applied Physics, Chinese Academy of Sciences, Shanghai 201800, China}
\affiliation{University of Chinese Academy of Sciences, Beijing 100049, China}
\affiliation{Institute of Frontier and Interdisciplinary Science \& Key Laboratory of Particle Physics and Particle Irradiation (MOE), Shandong University, Qingdao 266237, China}

 \author{Peng Huo}
 \affiliation{Department of Chemistry, Stony Brook University, Stony Brook, New York 11794-3800, USA}
 
\author{Jiangyong Jia}
 \affiliation{Department of Chemistry, Stony Brook University, Stony Brook, New York 11794-3800, USA}
 \affiliation{Physic Department, Brookhaven National Laboratory, Upton, New York 11796, USA} 

\author{Guo-Liang Ma}
\email[E-Mail:]{glma@sinap.ac.cn}
\affiliation{Shanghai Institute of Applied Physics, Chinese Academy of Sciences, Shanghai 201800, China}

%\email[]{glma@sinap.ac.cn}

%\date{\today}

\begin{abstract}
A new subevent cumulant method was recently developed, which can significantly reduce the non-flow contributions in long-range correlations for small systems compared to the standard cumulant method. In this work, we study multi-particle cumulants in $p$+Pb collisions at $\sqrt{s_{\mathrm{NN}}} = 5.02$ TeV with a multiphase transport model (AMPT), including two- and four-particle cumulants ($c_{2}\{2\}$ and $c_{2}\{4\}$) and symmetric cumulants [SC(2, 3) and SC(2, 4)]. Our numerical results show that $v_{2}\{2\}$ is consistent with the experimental data, while the magnitude of $c_{2}\{4\}$ is smaller than the experimental data, which may indicate either the collectivity is underestimated or some dynamical fluctuations are absent in the AMPT model. For the symmetric cumulants, we find that the results from the standard cumulant method are consistent with the experimental data, but those from the subevent cumulant method show different behaviors. The results indicate that the measurements from the standard cumulant method are contaminated by non-flow effects, especially when the number of produced particles is small. The subevent cumulant method is a better tool to explore the $real$ collectivity in small systems.
\end{abstract}


\maketitle

\section{Introduction}
\label{sec:intro}

One experimental signature suggesting the formation of nearly perfect fluid in ultrarelativistic nucleus-nucleus (A+A) collisions is the azimuthal anisotropy of produced particles. The measured anisotropies provide strong evidence of collective flow, which is commonly believed to be related to the hot QCD medium that expands collectively and transfers asymmetries in the initial geometry space into azimuthal anisotropies of produced particles in the final momentum space~\cite{Ollitrault:1992bk,Teaney:2000cw,Song:2007ux,Luzum:2008cw,Bozek:2009dw,Schenke:2010rr}. The feature of collectivity appears in the form of ``ridge'': enhanced pair production in a small azimuthal angle interval,  $\Delta\phi \sim 0$, extended over a wide range of pseudorapidity intervals $\Delta\eta$~\cite{Abelev:2009af,Alver:2009id,Chatrchyan:2011eka,ATLAS:2012at}. The azimuthal structure of the ridge is typically analyzed via a Fourier decomposition, $dN_{\mathrm{pairs}}/d\Delta\phi \sim 1+2\sum v_{n}^{2}\cos(n\Delta\phi)$. The second (elliptic; $v_{2}$) and third (triangular; $v_{3}$) Fourier harmonics are under intensive studies, because they are assumed to directly reflect the medium response to the initial geometry. For a small collision system, such as proton-proton ($p+p$) or proton-nucleus ($p$+A) collisions, it was assumed that the transverse size of the produced system is too small compared to the mean free path of constituents. Thus, it was expected that the collective flow in small systems should be much weaker than that in A+A collisions. However, recent observations of large long-range ridge-like correlations and $v_{n}$ coefficients in small systems~\cite{Khachatryan:2010gv, CMS:2012qk, Abelev:2012ola, Aad:2012gla, Adare:2013piz, Aad:2015gqa,Aad:2014lta} challenges the above paradigm of collective flow.

Since hydrodynamic flow implies a global collectivity involving all particles in the event, $k$-particle azimuthal cumulants, $c_n\{k\}$, are often used to measure the {\it true} $v_n$~\cite{Borghini:2000sa,Bilandzic:2010jr}. The standard cumulant method, known as the Q-cumulant~\cite{Bilandzic:2010jr}, use all $k$-particle multiplets in the entire detector acceptance to calculate $c_n\{k\}$.  But this method can be contaminated by non-flow effects, like jet-like correlation, especially when the multiplicity is small. Recently an improved cumulant method, referred to as the ``subevent cumulant,'' in which particles are divided into different subevents separated in the pseudorapidity $\eta$ direction, was developed~\cite{Jia:2017hbm}. Compared to the standard cumulant method, the new method can more effectively suppress intra-jet (single-jet) and inter-jet (di-jet) correlations. Recent ATLAS measurements have shown that the subevent method provides a more precise determination of $c_{n}\{4\}$ associated with long-range collectivity in small systems~\cite{Aaboud:2017blb}. 

Multi-particle correlation between different orders of flow harmonics is another complementary observable which provides additional constraints on the medium properties~\cite{Bilandzic:2013kga, ALICE:2016kpq}. Such mixed-harmonic correlations are measured through the so-called symmetric cumulant, SC(n, m), with $n \neq m$. The CMS Collaboration recently obtained results for SC(2, 3) and SC(2, 4) in $p+p$ and $p$+Pb collisions based on the standard cumulant method~\cite{Sirunyan:2017uyl}. However, Huo et al.  argued that the measurements of SC(n, m) in small systems are not trustworthy due to dominating non-flow effects, unless the subevent method is utilized~\cite{Huo:2017nms}. But their argument is based on the PYTHIA and HIJING models, which have no collective flow. Therefore it is necessary to verify this assertion with models that contain both collective flow and non-flow. 

Two classes of theoretical scenarios have been proposed to explain the collectivity in small sytems: hydrodynamical (or transport) models, which respond to initial geometry through final state interactions~\cite{Bozek:2011if,Bzdak:2013zma,Shuryak:2013ke,Qin:2013bha,Bozek:2013uha,Ma:2014pva,Bzdak:2014dia,Koop:2015wea,Shen:2016zpp,Weller:2017tsr,Song:2017wtw}, and the color glass condensate framework, which reflects the initial momentum correlation from gluon saturation effects~\cite{Dumitru:2010iy,Dusling:2013qoz,Skokov:2014tka,Schenke:2015aqa,Schlichting:2016sqo,Kovner:2016jfp,Iancu:2017fzn,Dusling:2015gta}. Both scenarios can describe the current experimental results. For example, a multiphase transport (AMPT) model with a tuned elastic parton-parton cross section $\sigma$= 3 mb can naturally reproduce the long-range two-particle azimuthal correlation and two-particle $v_n$ in high-multiplicity $p$+Pb and $p+p$ collisions and show good agreement with the experimental data~\cite{Ma:2014pva, Bzdak:2014dia}. However, the collectivity from the AMPT model has been interpreted as a parton escape mechanism where the azimuthal anisotropy is mainly generated by the anisotropic parton escape instead of hydro-like interactions~\cite{He:2015hfa,Lin:2015ucn}. The controversy surrounding the origin of collectivity in small systems needs to be further tested in more experimental and theoretical efforts. 

%To explore the $real$ collectivity in small systems, non-flow effects necessarily need to be suppressed. 
In this work, we adopt the newly developed subevent cumulant method to suppress non-flow effects to investigate the flow in the AMPT model for $p$+Pb collisions at $\sqrt{s_{NN}} = 5.02$ TeV. The two- and four-particle azimuthal cumulants, ($c_{2}\{2\}$ and $c_{2}\{4\}$),  and multi-particle azimuthal correlations between $v_2$ and $v_3$ and between $v_2$ and $v_4$, [SC(2, 3) and SC(2, 4)], are calculated using both standard and subevent cumulant methods. We find that the AMPT model can well describe the two-particle $v_{2}\{2\}$ data, but with a magnitude of $c_{2}\{4\}$ smaller than the experimental data. To further shed light on the origin and evolution of multi-particle correlations, the evolution of $c_{2}\{k\}$ values is traced at different phases in the AMPT model. Significant differences in symmetric cumulants, SC(2, 3) and SC(2, 4), between the standard and the subevent cumulant methods are also observed. Our results suggest that either the collectivity is  underestimated or some non-Gaussian dynamical fluctuations are missing in the AMPT model. We find that the subevent cumulant method is a better probe to investigate the $real$ collectivity in small systems.

\section{The AMPT Model}
\label{sec:model}
A multiphase transport model~\cite{Lin:2004en}, which is a hybrid dynamical transport model, is utilized in this work. We use the string melting AMPT version to simulate $p$+Pb collisions at $\sqrt{s_{\mathrm{NN}}}=5.02$ TeV. The string melting version consists of four main components: fluctuating initial conditions from the HIJING model~\cite{Wang:1991hta}, elastic parton cascade simulated by the ZPC model~\cite{Zhang:1997ej} for all partons from the melting of hadronic strings, a quark coalescence model for hadronization, and hadron rescatterings described by the ART model~\cite{Li:1995pra}. For details, see the review~\cite{Lin:2004en}. For the setting of parameter values,  we follow the recent AMPT study with a modest elastic parton-parton cross section $\sigma$ = 3 mb, which has been shown to be capable of reproducing the long-range correlation and two-particle $v_n$ coefficients in $p$+Pb collisions at $\sqrt{s_{\mathrm{NN}}} = 5.02$ TeV~\cite{Ma:2014pva, Bzdak:2014dia}.

\section{Multiparticle Cumulants}
\label{sec:Cumulant}
The cumulant method has been developed to characterize multi-particle correlations related to the collective expansion of system, while reducing non-flow contributions order by order~\cite{Borghini:2000sa, Borghini:2001vi}. A 2$k$-particle azimuthal correlator $\lr{\lr{2k}}$ is obtained by averaging over all unique combinations in one event then over all events, where the first two terms are $\lr{\lr{2}} =  \lr{\lr{ e^{in(\phi_{1}-\phi_{2}) }} }$ and $\lr{\lr{4}} =  \lr{\lr{ e^{in(\phi_{1}+\phi_{2} -\phi_{3}-\phi_{4}) }}}$. For a given harmonic $n$, the two- and four-particle cumulants can be determined:

\begin{equation}
\label{eq:c22}
c_{n}\{2\}  =  \lr{\lr{2}},
\end{equation}
\begin{equation}
\label{eq:c24}
c_{n}\{4\}  =  \lr{\lr{4}} - 2\lr{\lr{2}}^{2}.
\end{equation}

The flow coefficients $v_{n}$ can be analytically obtained from the two- and four-particle cumulants:
\begin{equation}
\label{eq:v22}
v_{n}\{2\}  =  \sqrt{c_{2}\{2\}},
\end{equation}
\begin{equation}
\label{eq:v24}
v_{n}\{4\}  =  \sqrt[4]{-c_{n}\{4\}}.
\end{equation}

The framework of the standard cumulant~\cite{Bilandzic:2010jr} expresses multi-particle correlations in terms of powers of the flow vector $Q_{n} = \sum e^{in\phi}$. The multi-particle correlations and cumulants can be calculated through a single loop over all events. In the standard cumulant method, the particles are chosen from the entire detector acceptance. In small systems, the non-flow correlations, especially the jet and dijet, dominate the azimuthal correlations. Hence, the standard cumulant may be strongly biased by these non-flow correlations, while the subevent cumulant method is designed to further suppress these non-flow correlations. In the subevent cumulant method, the entire event is divided into two subevents or three subevents.  Specifically, in the two-subevent method, the event is divided into two (labelled as $a$ and $b$) according to $-\eta_{\mathrm{max}} < \eta_{a} < 0$ and $0 < \eta_{b} < \eta_{\mathrm{max}}$; in the three-subevent method, the event is divided into three (labelled as $a$, $b$ and $c$) according to $-\eta_{\mathrm{max}} < \eta_{a} < -\eta_{\mathrm{max}}/3$, $-\eta_{\mathrm{max}}/3 < \eta_{b} < \eta_{\mathrm{max}}/3$ and $\eta_{\mathrm{max}}/3 < \eta_{c} < \eta_{\mathrm{max}}$. Then one can get the 2$k$-particle azimuthal correlators as follows:
\begin{equation}
\label{eq:2subevent}
\lr{\lr{2}} _{\mathrm{two-sub}}    =  \lr{\lr{ e^{in(\phi_{1}^{a}-\phi_{2}^{b}) }} },
\end{equation}

\begin{equation}
\label{eq:2subevent1}
\lr{\lr{4}}_{\mathrm{two-sub}}     =  \lr{\lr{ e^{in(\phi_{1}^{a}+\phi_{2}^{a} -\phi_{3}^{b}-\phi_{4}^{b}) }}},
\end{equation}

\begin{equation}
\label{eq:3subevent}
\lr{\lr{4}}_{\mathrm{three-sub}}   =  \lr{\lr{ e^{in(\phi_{1}^{a}+\phi_{2}^{a} -\phi_{3}^{b}-\phi_{4}^{c}) }}}.
\end{equation}

The symmetric cumulant is also based on the multi-particle cumulants, which measures the correlation between different flow harmonics on the basis of event-by-event fluctuations. The SC(n, m) is defined below:
\begin{equation}
\label{eq:SC}
SC(n, m)  =  \lr{\lr{ e^{in(\phi_{1}-\phi_{2}) + im(\phi_{3}-\phi_{4}) }}} - \lr{\lr{ e^{in(\phi_{1}-\phi_{2}) }} } \lr{\lr{ e^{im(\phi_{1}-\phi_{2}) }} } =   \lr{v_m^2v_n^2} -\lr{v_m^2}\lr{v_n^2}.
\end{equation}
Similarly, we can easily get the SC(n, m) with the subevent methods,

\begin{equation}
\label{eq:SC2sub}
SC(n, m)_{\mathrm{two-sub}}  =  \lr{\lr{ e^{in(\phi_{1}^{a}-\phi_{2}^{b}) + im(\phi_{3}^{a}-\phi_{4}^{b}) }}} -  \lr{\lr{ e^{in(\phi_{1}^{a}-\phi_{2}^{b}) }} } \lr{\lr{ e^{im(\phi_{1}^{a}-\phi_{2}^{b}) }} },
\end{equation}

\begin{equation}
\label{eq:SC3sub}
SC(n, m)_{\mathrm{three-sub}}  =   \lr{\lr{ e^{in(\phi_{1}^{a}-\phi_{2}^{b}) + im(\phi_{3}^{a}-\phi_{4}^{c}) }}} -  \lr{\lr{ e^{in(\phi_{1}^{a}-\phi_{2}^{b}) }} } \lr{\lr{ e^{im(\phi_{1}^{a}-\phi_{2}^{c}) }} }.
\end{equation}
More details can be found in Ref.~\cite{Jia:2017hbm}.

In order to make our results directly comparable to the experimental measurements, we choose $\eta_{\mathrm{max}}$ = 2.5 in our analysis to mimic the ATLAS detector acceptance for charged particles. The event selection is based on $\lr {N_{\mathrm{ch}} }$, the number of charged particles in $|\eta| < 2.5$ and $ p_{T} > 0.4$ GeV. The cumulant calculations are carried out using the charge particles in $|\eta| < 2.5$ and a certain $p_{T}$ selection, $0.3 < p_{T} <3$ GeV, and the number of charged particle in this $p_{T}$ range, $ N_{\mathrm{ch}}^{\mathrm{sel}}$. We need to point out $ N_{\mathrm{ch}}^{\mathrm{sel}} $ and $N_{\mathrm{ch}} $ are not the same due to different $p_{T}$ ranges. Then $\lr{2k}$ is averaged over events with the same $ N_{\mathrm{ch}}^{\mathrm{sel}} $ to obtain the $\lr{\lr{2k}}$ and SC(n, m). Finally, the cumulant results are obtained by mapping $ N_{\mathrm{ch}}^{\mathrm{sel}} $ to $\lr {N_{\mathrm{ch}}}$, where we follow the ATLAS procedure exactly with the same kinematic cuts~\cite{Aaboud:2017blb}.

\section{Results and Discussions}
\label{sec:results}

\begin{figure}
	\centering
	\includegraphics[scale=0.5]{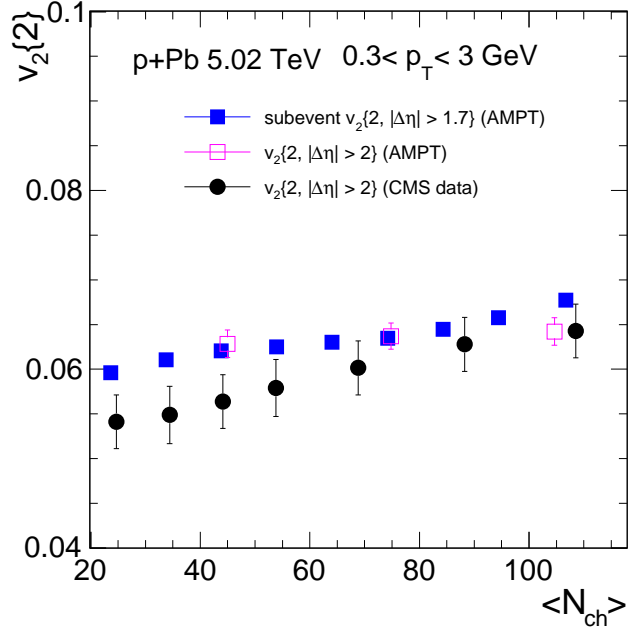}
	\caption{(Color online) $v_{2}\{2\}$ as a function of the number of charge particles $\langle N_{\mathrm{ch}}\rangle$ in $p$+Pb collisions at $\sqrt{s_{\mathrm{NN}}} = 5.02$ TeV, where filled squares represent the AMPT results using the subevent method, while open squares and filled circles represent the two-particle correlation results (with $|\Delta\eta| > 2 $) from the published AMPT results~\cite{Bzdak:2014dia} and CMS data~\cite{Chatrchyan:2013nka}, respectively. }
	\label{fig:fig1v22}
\end{figure}

Figure ~\ref{fig:fig1v22} shows the $v_{2}\{2\}$ results with the subevent cumulant method, and compares them with the two-particle $v_{2}\{2, |\Delta\eta| > 2\} $ from the published AMPT results~\cite{Bzdak:2014dia} and the CMS data~\cite{Chatrchyan:2013nka}.  The three results are in good agreement.

\begin{figure}
	\centering
	\includegraphics[scale=0.5]{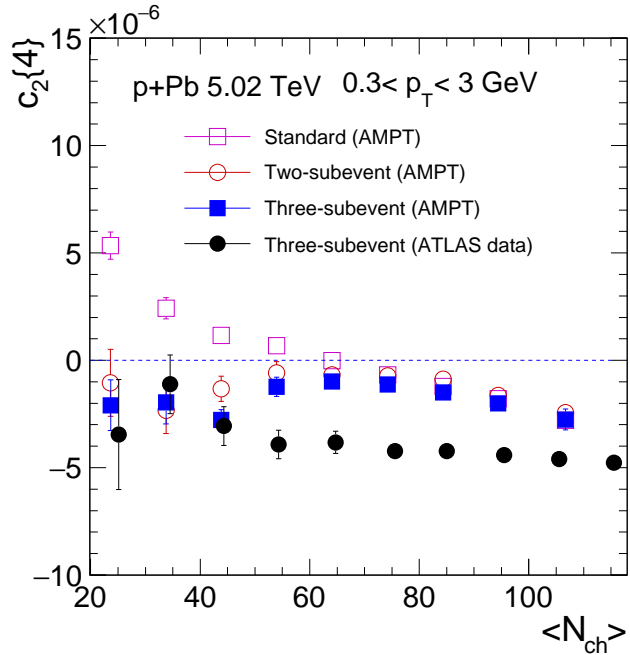}
	\caption{(Color online) $c_{2}\{4\}$ as a function of the number of charge particles $\langle N_{\mathrm{ch}}\rangle$ in $p$+Pb collisions at $\sqrt{s_{\mathrm{NN}}} = 5.02$ TeV, where open squares, open circles, and filled squares represent the AMPT results using the standard cumulant, two-subevent, and three-subevent methods, respectively. The filled circles represent the ATLAS data using three-subevent cumulant method~\cite{Aaboud:2017blb}.}
	\label{fig:fig2c24}
\end{figure}

Figure~\ref{fig:fig2c24} shows the $c_{2}\{4\}$ results as a function of the number of charge particles, where the AMPT results are calculated with three methods, (standard cumulant, two-subevent cumulant and three-subevent cumulant methods), in comparison with the experimental data. We find that $c_{2}\{4\}$  using the standard cumulant method is negative at $\lr {N_{\mathrm{ch}} } > 70$, but changes to positive at $\lr {N_{\mathrm{ch}} } < 70$, a region expected to be more affected by non-flow contributions. In contrast to the standard cumulant method, the $c_{2}\{4\}$ from subevent methods remains negative over the full $\lr {N_{\mathrm{ch}} }$ region. It is surprising that the $c_{2}\{4\}$ using two-subevent method agrees almost completely with that using the three-subevent method. This may indicate that the two-subevent method already suppresses most of the non-flow contributions in the AMPT model. On the other hand, the magnitude of $c_{2}\{4\}$ with the AMPT model is systematically smaller than that with the ATLAS data. Since $c_{2}\{4\}$ is sensitive to not only the averaged collectivity $\lr{v_2}$ but also the shape of the $v_2$ probability distribution $p(v_2)$~\cite{Jia:2014pza,Yan:2013laa} ~\cite{Voloshin:2007pc}, the lack of $c_{2}\{4\}$ may indicate that either the collectivity is underestimated or some non-Gaussian dynamical fluctuations of $v_2$ are missing in the AMPT model~\cite{Ma:2014xfa}.

\begin{figure}
	\centering
	\includegraphics[scale=0.5]{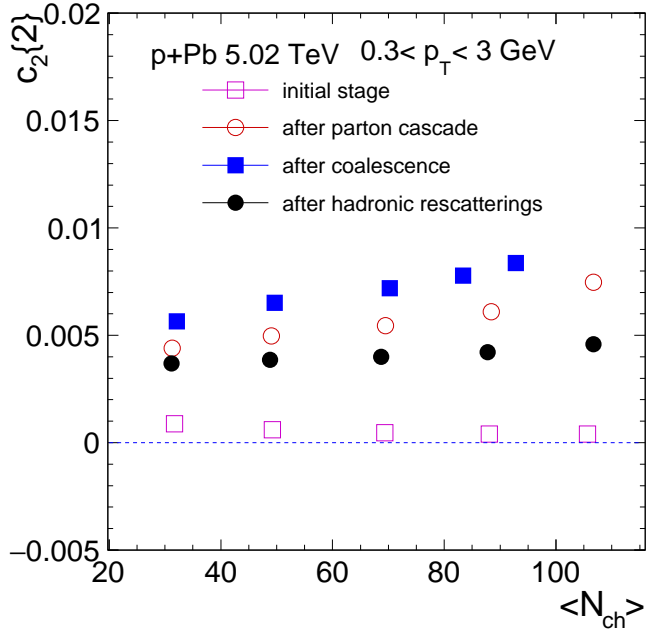}
	\caption{(Color online) AMPT results on $c_{2}\{2\}$ with the three-subevent method as a function of the number of charge particles  $\langle N_{\mathrm{ch}}\rangle$ for four different evolution stages, i.e., initial stage (open squares), after parton cascade (open circles), after coalescence (filled squares), and after hadronic recatterings (filled circles), in $p$+Pb collisions at $\sqrt{s_{\mathrm{NN}}} = 5.02$ TeV.}
	\label{fig:fig3timeC22}
\end{figure}

\begin{figure}
	\centering
	\includegraphics[scale=0.5]{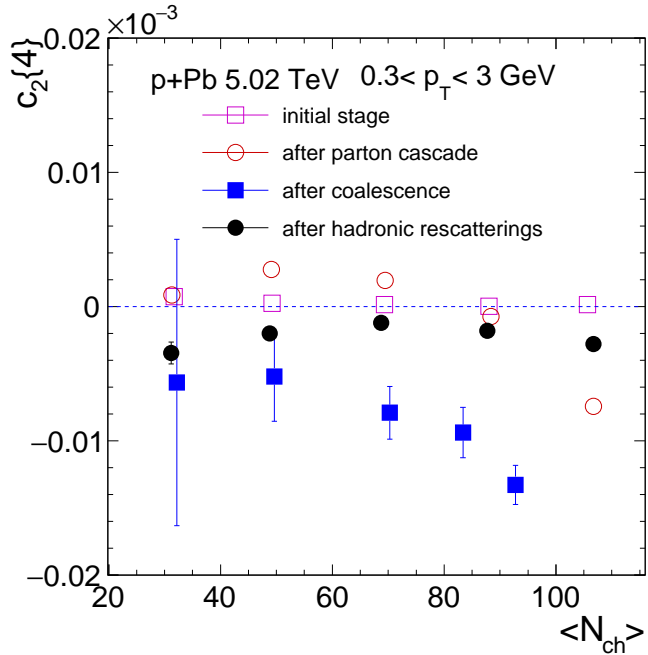}	
	\caption{Same as Figure~\ref{fig:fig3timeC22} but for $c_{2}\{4\}$ with the three-subevent method.}
	\label{fig:fig3timeC24}
\end{figure}

To further investigate the collectivity behavior, we trace the values of $c_{2}\{2\}$ and $c_{2}\{4\}$ at four different evolution stages: initial stage, after parton cascade, after coalescence and after hadronic rescatterings. The $c_{2}\{2\}$ and $c_{2}\{4\}$ at these four stages are all calculated using the three-subevent method, as shown in Figures~\ref{fig:fig3timeC22} and ~\ref{fig:fig3timeC24}, respectively. At the initial stage, the $c_{2}\{2\}$ and $c_{2}\{4\}$ are slightly positive at small $\lr {N_{\mathrm{ch}} }$ and asymptotically approach zero towards larger $\lr {N_{\mathrm{ch}} }$. This behavior is consistent with the expectation from transverse momentum conservation~\cite{Bzdak:2017zok}. After parton cascade, $c_{2}\{2\}$ is enhanced and $c_{2}\{4\}$ changes sign from positive to negative at a certain value of $\lr {N_{\mathrm{ch}}}$, which maybe due to the interplay between transverse momentum conservation and an anisotropic flow generated by parton cascade~\cite{Bzdak:2018web}. After coalescence, the more positive $c_{2}\{2\}$ and more negative $c_{2}\{4\}$ are seen for all charged hadrons not including resonances, which indicates that the strength of collective correlations increases, since both $\lr {p_{T} }$ and $v_{2}(p_T)$ can be enhanced via the coalescence process. In the final stage, i.e., after hadronic rescatterings, the magnitudes of $c_{2}\{2\}$ and $c_{2}\{4\}$ decrease significantly because of the cooling down of systems. However, we notice that hadronic rescatterings suppress $c_{2}\{4\}$ more strongly than $c_{2}\{2\}$ (by a factor of $\sim$ 8 vs 2); the detailed dynamics of this behavior deserves further investigation in the future.

\begin{figure}
	\centering
	\includegraphics[scale=0.5]{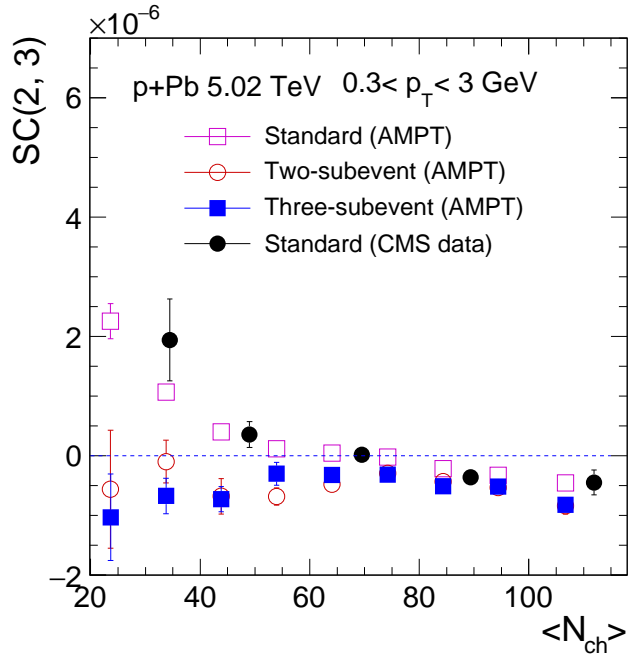}
	\caption{(Color online) SC(2, 3) as a function of the number of charge particles  $\langle N_{\mathrm{ch}}\rangle$ from AMPT calculations using the standard cumulant (open squares), two-subevent (open circles) and three-subevent (filled squares) methods, in comparison with the experimental data with standard cumulant method (filled circles)~\cite{Sirunyan:2017uyl}, in $p$+Pb collisions at $\sqrt{s_{\mathrm{NN}}} = 5.02$ TeV.}
	\label{fig:fig4sc23}
\end{figure}

\begin{figure}
	\centering
	\includegraphics[scale=0.5]{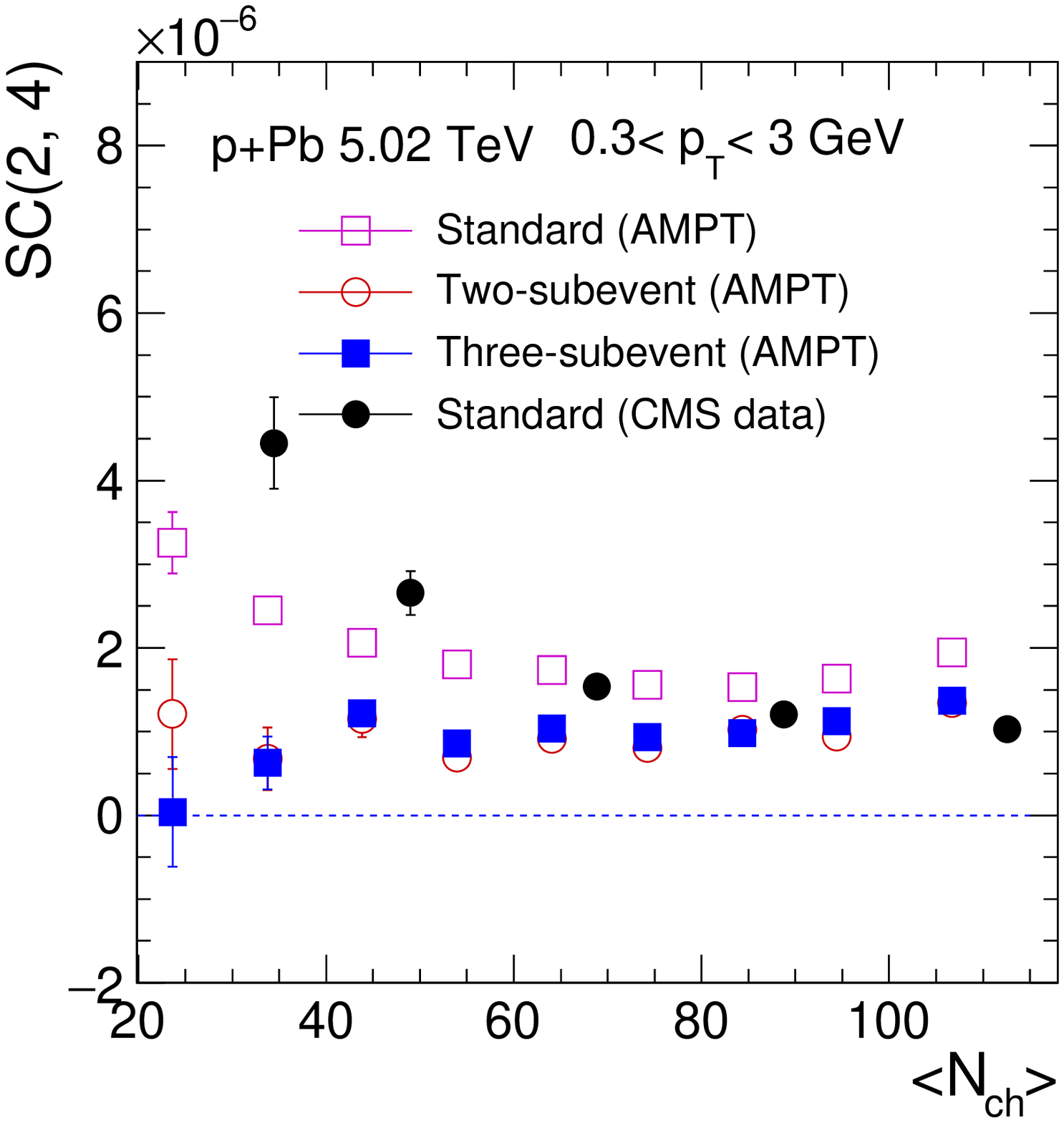}
	\caption{Same as Figure~\ref{fig:fig4sc23} but for SC(2, 4).}
	\label{fig:fig5sc24}
\end{figure}

Figures~\ref{fig:fig4sc23} and ~\ref{fig:fig5sc24} show the symmetric cumulants SC(2, 3) and  SC(2, 4), respectively. The results using three methods are presented (standard cumulant, two-subevent cumulant and three-subevent cumulant methods), in comparison with the CMS data which are based on the standard cumulant method. We find that the SC(2, 3) from the standard cumulant method is negative at a high multiplicity, while it becomes positive at low multiplicity, which is in good agreement with the CMS data. However, we find the SC(2, 3) from the subevent methods stays negative for the whole range of multiplicity. Our results strongly suggest that the measurements using the standard cumulant method are contaminated by the non-flow effects. On the other hand, the SC(2, 4) from the standard cumulant method is comparable with the experimental data, but it is much larger than those with subevent methods. It also suggests that non-flow contributions need to be removed to obtain a clean signal of collectivity, especially in the low-multiplicity region in small systems.

\section{CONCLUSIONS}
\label{sec:conclusion}
The subevent cumulant method is utilized to study multi-particle correlations in $p$+Pb collisions within the AMPT model. The two- and four-particle cumulants, ($c_{2}\{2\}$ and $c_{2}\{4\}$), and multi-particle azimuthal correlations between different flow harmonics, [SC(2, 3) and SC(2, 4)], are numerically calculated. The $v_{2}\{2\}$ is consistent with the experimental data, while the magnitude of $c_{2}\{4\}$ is systematically smaller than the experimental data. This behavior indicates that either the collectivity is underestimated or some non-Gaussian dynamical fluctuations are absent in the AMPT model. The SC(2, 3) from the standard cumulant method is negative at a high multiplicity, but changes sign towards a low multiplicity. However, the SC(2, 3) from the subevent cumulant method is negative for the whole range of multiplicities. The SC(2, 4) from the standard cumulant method is larger than those from the subevent cumulant methods. These results suggest that the measurements based on the standard cumulant method are contaminated by the non-flow effects, and the subevent cumulant method should be used to investigate the $real$ collectivity in small systems.

\section*{ACKNOWLEDGMENTS}

M.-W. N. and G.-L. M are supported by the National Natural Science Foundation of China under Grant Nos. 11522547, 11375251, and 11421505, and the Major State Basic Research Development Program in China under Grant No. 2014CB845404. J.J. and P.H. are supported by the National Science Foundation under Grant No. PHY-1613294.

\end{document}